\documentclass[english,keywords,aps,amsmath,amssymb,twocolumn]{revtex4-1}
\usepackage[T1]{fontenc}
\usepackage[latin1]{inputenc}
\usepackage{babel}
\usepackage{amssymb}
\usepackage{graphicx}
\usepackage{color}
\usepackage{bm}
\usepackage{longtable}
\usepackage{amsmath} 
\usepackage{amsfonts}
\usepackage{amssymb}
\begin{document}
\title{ On the violation of L$\ddot{u}$ders bound of macrorealist and noncontextual inequalities}
\author{Asmita Kumari\footnote{asmita.physics@gmail.com }}
\author{Md. Qutubuddin\footnote{qutubuddinsxc@gmail.com }}
\author{A. K. Pan \footnote{akp@nitp.ac.in}}
\affiliation{National Institute of Technology Patna, Ashok Rajpath, Patna, Bihar 800005, India}
\begin{abstract}
In a recent Letter [PRL, 113, 050401 (2014)], it is shown that the quantum violation of a three-time Leggett-Garg inequality (LGI) for a dichotomic qutrit system can exceed the L$\ddot{u}$ders bound. This is obtained by using a degeneracy breaking projective measurement rule which the authors termed as von Neumann rule. Such violation can even approach the algebraic maximum in the asymptotic limit of system size. In this paper, we question the implication of such violation of L$\ddot{u}$ders bound and its conceptual relevance in LG scenario. We note an important fact that the basis for implementing the proposed von Neumann rule for a degenerate observable is non-unique  and show that the violation of L$\ddot{u}$ders bound is crucially dependent on the choice of basis. Further, we demonstrate  the violation of L$\ddot{u}$ders bound of the simplest non-contextual inequality (NCI) which is in contrast to the reasoning provided in the aforementioned Letter. This result further raises the doubts regarding  the validity of the proposed rule as a viable projective measurement. We discuss the relevance of such results with respect to the usual quantum violation of LGI and NCI. 

\end{abstract}

\maketitle

\section{Introduction}

In their epoch-making paper, Einstein, Podolsky and Rosen \cite{epr35} had raised the fundamental question about the incompleteness of quantum mechanical description of nature by the $\psi$-function. The realist hidden variable models are those which assume to provide the `complete state' of the system with the aid of so-called hidden variables along with the quantum state. In this regard, the central question is what constraint a realist model of quantum phenomena has to satisfy  to be compatible with the empirically verifiable predictions of quantum mechanics (QM). In his pioneering work in 1964, Bell \cite{bell64} first pointed out that any realist model reproducing the QM has to be nonlocal. Since then, extensive studies (see, for reviews, \cite{genovese,rev}) have been performed for revealing the nonlocal character of QM both theoretically and experimentally.

Another constraint, known as contextuality, was discovered by Kochen and Specker  \cite{ks67}. According to  the assumption of non-contextuality in a realist hidden variable model, the definite value assignment to a measurement is independent of the compatible measurements that are being performed jointly or sequentially. Kochen and Specker theorem demonstrates that such a value assignment is in contradiction with the statistics of QM for a certain set of observables. The original proof of the Kochen and Specker theorem involves a complex structure using 117 vectors. Later, simplified versions have been proposed by Peres \cite{peres} and Mermin \cite{mermin90}. These proofs are converted into testable inequalities, valid for any non-contextual realist model, but violated by QM \cite{cab08,pan10}. The notion of contextuality has also been extensively studied both theoretically \cite{theocontext,kle} and experimentally \cite{expcontext}.

Of late, the macrorealist models and its compatibility with the QM has also been attracting increasing attention. Such a line of study was first put forwarded by Leggett and Garg \cite{lg85} by introducing a set of inequalities (henceforth, LGIs) for testing the compatibility between the classical world view of macrorealism and QM. These inequalities play similar role to that of the Bell inequalities in testing local hidden-variable models but involving a single system subjected to the measurements at different times. Leggett and Garg inequalities are derived by considering the assumptions of \textit{macrorealism per se} and \textit{non-invasive measurability} . According to the assumption of macrorealism \emph{per se}, at any instant of time a system remains in one of it's macroscopically distinct ontic state, whereas the non-invasive measurability ensures that the ontic state of the system remains uninfluenced by the measurement and dynamics \cite{lg85,lagtt2002}. In recent times, a flurry of theoretical studies have been reported \cite{emary12,maroney14,kofler13,budroni15,budroni14,emary,halliwell16,kofler08,saha15,swati17,pan17} and a number of experiments have been conducted by using different systems \cite{lambert,goggin11,knee12,laloy10,george13,knee16,kati1,wang02}.

The quantum violations of local, non-contextual and macrorealist inequalities  are witnesses of  non-classicality. Improved quantum violation of a given inequality is thus the signature of more non-classicality. The present paper questions the issue of improved quantum violation of LGIs that is proposed in a recent work \cite{budroni14}. The simplest Bell's inequality is the CHSH one \cite{chsh69} whose quantum violation is restricted by the  Cirelson's bound \cite{cri}. LGIs are often considered structurally analogous to Bell's inequalities. But, Budroni and Emary \cite{budroni14} have recently shown that the former can be violated up to its algebraic maximum.  Specifically, by considering a  degeneracy breaking projective measurement scenario termed as  `von Neumann rule', they showed that the quantum violation of the LGI can exceed the  L$\ddot{u}$ders bound and can even approach the algebraic maximum at the asymptotic limit of the system size. Such amount of violation of CHSH inequalities may be achieved in post-quantum theories but not in QM. For dichotomic observables, the maximum quantum violation of a three-time LGI  is $1.5$ if L$\ddot{u}$ders projection rule \cite{lu} is used, irrespective of the system size \cite{budroni13}. Authors in \cite{budroni14} have shown that for dichotomic observables in a qutrit system and for suitable choice of initial quantum state, the use of such a degeneracy breaking rule provides the quantum violation of LGI $1.75$ exceeding the L$\ddot{u}$ders bound. Using semi-definite programming they have shown that such a violation can reach 2.21. Such a result is verified in recent  experiment \cite{kati1,wang02} using qutrit system. Additionally, they remarked that the L$\ddot{u}$ders bound of a non-contextual inequality (NCI) \emph{cannot} be violated in this manner. In support of their claim, they mentioned that since in contextuality test the joint measurements are assumed for commuting observables, then which state update rule is being used becomes irrelevant even when the measurements are performed sequentially \cite{budroni14}. 

Before proceeding further, let us first encapsulate the essence of L$\ddot{u}$ders rule and the von Neumann rule proposed in \cite{budroni14}. Consider an observable $\hat{A}$ having discrete eigenvalues $a_1, a_2, a_3$... $a_m$ with degree of degeneracies $x_1,x_2,x_3$...$x_m$ respectively. Let  $P_{m}^{\alpha}=|\phi_{m}^{\alpha}\rangle\langle\phi_{m}^{\alpha}|$ is the projection operator associated with $m^{th}$ eigenvalue  where $\alpha$ denotes the degeneracy. The von Neumann projection rule breaks the degeneracy, so that, the reduced density matrix can be written as $\rho_{v}=\sum_{m,\alpha}{P_{m}^{\alpha}\rho P_{m}^{\alpha}}$ where   $\rho$ is the initial density matrix of the system. As already indicated, $\rho_{v}$ is \textit{not} unique for degenerate observable. On the other hand, the L$\ddot{u}$ders projection rule respects the degeneracy. The reduced density matrix in this case can be written as $\rho_{l}=\sum_{m}{P_{m}\rho P_{m}}$ where $P_{m}=\sum_{\alpha=1}^{x_{m}}|\phi_{m}^{\alpha}\rangle\langle\phi_{m}^{\alpha}|$  \cite{heger, pan}. Then, for an observable with degenerate eigenvalues, the von Neumann rule provides the reduced density matrix less coherent than that is obtained using the L$\ddot{u}$ders rule. For non-degenerate observable both the rules are identical. Throughout our paper by von Neumann rule we refer the discussion in this paragraph to avoid any confusion.

In this paper, we critically examine the reason of such a violation of L$\ddot{u}$ders bound of LGIs and question the validity of the so-called von Neumann rule in LG scenario. We first point out that the choice of basis for applying that rule is not unique. There can be infinitely many choices of basis for implementing it, even for a dichotomic observable in qutrit system. Thus, in a sequential measurements of two degenerate observables, the reduced state after the first measurement becomes non-unique which may then produce different results of sequential measurements of the same two observables. We indeed show that how different choice of von Neumann basis provides different amount of quantum violation of a given LGI. We provide some detail calculations to show how for a given state and observable the choice of von Neumann basis provide different amount of violations of LGIs.

Further, we study the quantum violation of non-contextual inequality (NCI). As opposed to the claim in \cite{budroni14}, we show that the use of the von Neumann rule violates the L$\ddot{u}$ders bound of NCI.  Surprisingly, we consider the simplest NCI involving three commuting dichotomic observables pertaining to a qutrit system. In such choices of observables the violation of L$\ddot{u}$ders bound of NCI is not expected. Similar to the case of LGIs, we find different choices of von Neumann basis that provide different amounts of quantum violations of a given NCI and a suitable choice of basis provide the quantum violation of NCI 2. 

A LGI consists of statistical correlations of sequential measurements of two observables. It seems thought-provoking why the violation of L$\ddot{u}$ders bound occurs when one uses von Neumann reduction rule.  To understand this issue let us consider a degenerate observable $A_1$. The measurement of which naturally uses L$\ddot{u}$ders projection rule. In order to forcefully implement the von Neumann projection rule \cite{heger}, one needs to perform a non-detective measurement  of another observable (say, $A_{1}^{\prime}$) prior to $A_1$, where $A_1$ and $A_{1}^{\prime}$ are commuting. Hence, the measured statistics of $A_{1}$ remains same irrespective of the fact whether it is measured solely or after $A_{1}^{\prime}$. However, the crucial fact is that the reduced density matrices for those two aforementioned measurement scenarios are different whose effect can be detectable if a subsequent measurement of another observable, say $A_2$ (may or may not be commuting with $A_1$), is performed separately on the two reduced density matrices.  Then, the statistics of such a sequential measurement can be different for two different density matrices because $A_{1}^{\prime}$ may not be commuting with $A_2$ in general. Moreover, different von Neumann basis produce different reduced density matrices. Similar argument holds good for other two sequential measurements involving the LGI. 

It is then intuitively clear that a kind of additional non-classicality is being introduced through the von Neumann projection rule which provides the violation of L$\ddot{u}$ders bound of LGIs and also of NCIs. In other words, one may say that it is a kind of quantum contextuality induced violation of  L$\ddot{u}$ders bound of LGIs and NCIs which cannot be considered as the violation of LGIs and NCIs in its usual sense. Note that, the prior observable $A_{1}^{\prime}$ is nowhere included in deriving the classical bounds of LGIs or NCIs. The inclusion of $A_{1}^{\prime}$s in the realist model formulation may provide a different classical bound of the relevant inequalities. Then, it is certainly erroneous to treat the violation of L$\ddot{u}$ders bound by von Neumann projection rule in the same footing as the quantum violation of usual LGIs and NCIs. 

This paper is organized as follows. In Sec.II, we show the difference between L$\ddot{u}$ders and von Neumann rules in sequential measurement of two degenerate qutrit observables and in Sec.III, we first show that given a particular von Neumann basis, the suitable intermediate evolutions can improve the maximum quantum violations of LGIs for a given state and observable and then we explicitly demonstrate that the violation of  L$\ddot{u}$ders bound is dependent on the choices of von Neumann basis. In Sec. IV, we demonstrate that quantum violations of NCIs exceeding the  L$\ddot{u}$ders bound if von Neumann projection rule is used. We discuss the implications of our study in Sec. V.

\section{ L$\ddot{u}$ders rule, von Neumann rule and Sequential measurement}

For a qutrit system, let us now assume two dichotomic degenerate observables $\hat{A}$ and $\hat{B}$ such that $P^{\alpha}_{m}$ and $P^{\beta}_{n}$ are their respective projectors. Here $\alpha$, $\beta$ are degeneracy and $m,n=\pm$ are the eigenvalues of $\hat{A}$ and $\hat{B}$ respectively. For the particular case of dichotomic observable in a qutrit system $\alpha,\beta=1,2$. Then the sequential measurement $\hat{A}$ and $\hat{B}$ provides
\begin{eqnarray}
{\langle \hat{A} \hat{B}\rangle}_{seq} &=& \sum_{m,n=\pm {1}}mn P(P^{\alpha}_{m}P^{\beta}_{n}) 
\end{eqnarray}
For L$\ddot{u}$ders rule the joint probability $P(P^{\alpha}_{m}P^{\beta}_{n})= Tr[{P_{{m}}}\rho P_{m}P_{n}]$ where $P_{m}=\sum_{\alpha}P^{\alpha}_{{m}}$, $P_{n}=\sum_{\beta}P^{\beta}_{{n}}$. On the other hand, using von Neumann rule the joint probability is given by $P(P^{\alpha}_{m}P^{\beta}_{n})= Tr[\sum_{\alpha,\beta}{P^{\alpha}_{{m}}} \rho P^{\alpha}_{m} P^{\beta}_{n}]$.

The sequential measurement by using L$\ddot{u}$ders rule provides 
\begin{eqnarray}
{\langle \hat{A} \hat{B}\rangle}^{L}_{seq} &=& \frac{1}{2}(Tr[\rho \hat{A}\hat{B}] +Tr[\rho \hat{B}\hat{A}])
\end{eqnarray}
which is irrespective of the dimension of the system. But, using von Neumann projection rule, for example, in the case of dichotomic observable in qutrit system,  we obtain 
\begin{eqnarray}
\label{lv}
{\langle \hat{A} \hat{B}\rangle}^{V}_{seq} = {\langle \hat{A} \hat{B}\rangle}^{L}_{seq} - Tr[(P^{\alpha_{1}}_{+}\rho P^{\alpha_{2}}_{+} + P^{\alpha_{2}}_{+}\rho P^{\alpha_{1}}_{+})\hat{B}]
\end{eqnarray} 
Here $P^{\alpha_{1}}_{+}$ and $P^{\alpha_2}_{+}$ are the projectors with same eigenvalue $m=+1$. It is then evident from Eq.(\ref{lv}) that there is an extra term along with ${\langle \hat{A} \hat{B}\rangle}^{L}_{seq}$ which is in general non-zero and responsible for the violation of  L$\ddot{u}$ders bound. Importantly, while ${\langle \hat{A} \hat{B}\rangle}^{L}_{seq}$  is basis independent, the quantity $Tr[(P^{\alpha_{1}}_{+}\rho P^{\alpha_{2}}_{+} + P^{\alpha_{2}}_{+}\rho P^{\alpha_{1}}_{+})\hat{B}]$ is dependent on the choice of von Neumann basis leading to the basis dependent violation of L$\ddot{u}$ders bound. We show how the choice of basis provides different amount of violation of LGIs and NCIs. 
 
\section{Violation of L$\ddot{u}$ders  bound of LGIs}
Let us assume a suitable dichotomic observable $\hat{M_1}$ at $t=t_1$, and its measurement on the state of a macroscopic system produce a definite outcome $1$ or $-1$ at any instant of time, as per the assumption of macrorealism \emph{per se}. In LG scenario, the measurement of $\hat{M_1}$ is performed on macroscopic system at three different times $t_1,t_2$ and $t_3$ ($t_3 > t_2 > t_1$) leads the measurement  observables $\hat{M_1}$,$\hat{M_2}$ and $\hat{M_3}$. The notion of non-invasive measurability ensure the existence joint probability of different outcomes $P(M_{1}^{\pm}M_{2}^{\pm}M_{3}^{\pm})$ and uninfluenced marginal effect on prior and future measurements.

Based on the aforementioned two assumptions, the LGIs are derived as
\begin{eqnarray}
\label{lgi1}
K_{13}=\langle \hat{M_{1}} \hat{M_{2}}\rangle + \langle \hat{M_{2}} \hat{M_{3}}\rangle -\langle \hat{M_{1}} \hat{M_{3}}\rangle \leq 1
\end{eqnarray}
\begin{eqnarray}
\label{lgi2}
K_{23} =\langle \hat{M_{1}} \hat{M_{2}}\rangle - \langle \hat{M_{2}} \hat{M_{3}}\rangle + \langle \hat{M_{1}} \hat{M_{3}}\rangle \leq 1
\end{eqnarray}
\begin{eqnarray}
\label{lgi3}
K_{12} =-\langle \hat{M_{1}} \hat{M_{2}}\rangle + \langle \hat{M_{2}} \hat{M_{3}}\rangle + \langle \hat{M_{1}} \hat{M_{3}}\rangle \leq 1
\end{eqnarray}
where $\langle \hat{M_{r}} \hat{M_{s}}\rangle$ is a correlation function of dichotomous observable $\hat{M_{r}}$ and $\hat{M_{s}}$, where $r,s = 1,2$ and $3$ with $r < s$. In QM, the $\hat{M_{r}}$ and  $\hat{M_{s}}$ are unitarily connected as $\hat{M_{s}}=U_{\Delta{t}_{rs}} \hat{M_{r}} U^{\dagger}_{\Delta{t}_{rs}}$, where $U_{\Delta{t}_{rs}} =e^{i H (t_{r} - t_{s}) }$. For a quantum state $\rho_{t_1}$ the correlation functions  $\langle \hat{M_{r}} \hat{M_{s}}\rangle = \sum_{m,n=\pm {1}}mn P(M^{m}_r M^{n}_s) $ where the joint probability can be obtained by using L$\ddot{u}$ders rule is $P(M^{m}_r M^{n}_s) = Tr[{U_{\Delta{t_{rs}}}{P_{{m}}} {U_{\Delta{t_{1r}}}}\rho_{t_1}U^{\dagger}_{\Delta{t_{1r}}}P_{m}U^{\dagger}_{\Delta{t_{rs}}}P_{n}}]$. Here $P_{m}=\sum_{\mu}\Pi^{\mu}_{{m}}$, $P_{n}=\sum_{\nu}\Pi^{\nu}_{{n}}$ and $\mu, \nu$ are the relevant degeneracies.
On the other hand, joint probability for the von Neumann rule is $P(M^{m}_r M^{n}_s) = Tr[\sum_{\mu,\nu}{U_{\Delta{t_{rs}}}{\Pi_{\mu}^{{m}}} {U_{\Delta{t_{1r}}}}\rho_{t_1}U^{\dagger}_{\Delta{t_{1r}}}\Pi_{\mu}^{{m}}U^{\dagger}_{\Delta{t_{rs}}}\Pi_{\nu}^{n}}]$. 

We consider the same observable $\hat {M_1}= |{3}\rangle\langle{3}| + |{2}\rangle\langle{2}| - |{1}\rangle \langle{1}|$ as  in \cite{budroni14}, where $|{1}\rangle=(1,0,0)^T ,|{2}\rangle=(0,1,0)^T$ and $|{3}\rangle=(0,0,1)^T$ are the eigenvectors with eigenvalues ($-1,1,1$) respectively. However, $\hat {M_1}$ can also be decomposed as $\hat {M_1}= |{3'}\rangle\langle{3'}| + |{2'}\rangle\langle{2'}| - |{1}\rangle \langle{1}|$ where $|{1}\rangle=(1,0,0)$, $|{2'}\rangle=\xi |{2}\rangle+ \sqrt{1-\xi^{2}}|{3}\rangle$ and $|{3'}\rangle=\sqrt{1-\xi^{2}}|{2}\rangle-\xi |{3}\rangle)$ are also eigenvectors having eigenvalues $(-1,1,1)$ respectively. For $\xi = 1$, former decomposition can be recovered which is the case considered by Budroni and Emary \cite{budroni14}. Importantly, L$\ddot{u}$ders measurement is independent of $\xi$ but it plays a crucial role in von Neumann rule as seen from Eq.(\ref{lv}). Different values of $\xi$ thus implement different von Neumann measurements. 

The evolution is governed by the Hamiltonian  $\label{ham.} H= \gamma \hat{J_{x}}$, where $\gamma$ the coupling constant and $\hat{J_{x}}$ is the angular momentum operator along $\hat{x}$ direction.  Then the unitary evolution can be written as  $U_{\Delta{t}_{rs}} = e^{i \gamma \hat{J_{x}} {\Delta{t}}_{rs}}$. We denote $g_1 =\gamma\Delta{t}_{12}$ and $g_2 =\gamma\Delta{t}_{23}$. Usually, in LG scenario the coupling strengths are taken to be same, i.e., $g_1=g_2$. Here we take $g_1\neq g_2$ which further improves the amount of maximum quantum violation. For qubit case different values of $g_1$ and $g_2$ do not improve the maximum quantum violation.  

Before proceeding further, let us first briefly recapitulate the essence of results in \cite{budroni14}. By considering the same state $|\psi_{t_1} \rangle=(0,0,1)^{T}$, the quantum mechanical expression of $K_{13}$ (say, $K_{13b}^{v}$) using von Neumann basis with $\xi = 1$ and   $g_1 = g_2=g$ derived by them is given by \cite{budroni14}
\begin{align}
\label{b1}
K^{v}_{13b} =\frac{1}{16}\big[ 1 + 32\cos(g) - 20 \cos(2 g)+3 \cos(4 g)\big]
\end{align}
The maximum quantum value of Eq. (\ref{b1}) is $1.75$ at $g = 1.31$, clearly exceeding the L${\ddot{u}}$ders bound $1.5$. They \cite{budroni14} have further showed that the violation increases with the increment of system size and approaches algebraic maximum $3$ in asymptotic limit of system size. 
\begin{table}[]
\begin{center}
\begin{tabular} {|p{2cm}|p{1.5cm}|p{1.5cm}|p{1.5cm}|}  
\hline
\multicolumn{4}{|c|}{Quantum violations of various LGIs for $\xi =1 $ } \\
\hline
\centering LGI & \centering Max. Value & \centering $g_1$ & \ \ \ \ \  $g_2$ \\
\hline
\centering $ K^{l}_{13}$ & \centering $1.45$ & \centering $\frac{\pi}{2}$ & \  \ \ \ \ $\frac{\pi}{4}$  \\
\centering $ K^{v}_{13b}$ & \centering $1.75$ & \centering $1.31$ & \ \ \ \ \ $1.31$ \\
\centering $K^{v}_{13}$ & \centering $1.91$ & \centering $0.98$ & \ \ \ \ \ $1.85$  \\
\hline
\hline
\centering $K^{l}_{23}$ & \centering $1$ & \centering $\pi$ & \ \ \ \ \ $\pi$ \\
\centering $K^{v}_{23b}$ & \centering $1$ & \centering $\pi$ & \ \ \ \ \ $\pi$ \\
\centering $ K^{v}_{23}$ & \centering $1.78$ & \centering $-\pi/3$ & \ \ \ \ \ $2\pi/3$ \\
\hline
\hline
\centering $K^{l}_{12}$ & \centering $1.45$ & \centering $\frac{3\pi}{4}$ & \ \ \ \ \ $\frac{-\pi}{4}$ \\
\centering $K^{v}_{12b}$ & \centering $1$ & \centering $\pi$ & \ \ \ \ \ $\pi$  \\
\centering $K^{v}_{12}$ & \centering $1.44$ & \centering $2.41$ & \ \ \ \ \ $-0.73$ \\
\hline
\end{tabular}
\end{center}
\caption{The quantities $K^{l}_{13}$, $K^{l}_{23}$ and $K^{l}_{12}$ denote the maximum values when L$\ddot{u}$ders measurement is performed, and $K^{v}_{13b}$, $K^{v}_{23b}$ and $K^{v}_{12b}$ are the values when von Neumann measurement with  $\xi = 1$ is performed and $g_{1} = g_{2}$ as used in \cite{budroni14}. $K^{v}_{13}$, $K^{v}_{23}$ and $K^{v}_{12}$ are calculated for $\xi =1$ but $g_{1} \neq g_{2}$. }
\end{table}

Now, instead of same coupling constant, if we take $g_1 \neq g_2$, we have 
\begin{eqnarray}
\label{b2}
K^{v}_{13}&=& \frac{1}{2}\Big[\sin^2({g_1}) + \cos({g_2})+2 \cos^2({g_1})\cos({g_2})\nonumber\\& +&  \cos^2({g_2})\sin^2({g_1}) + \cos({g_1})\left(2 + \sin^2({g_2})\right)\nonumber\\& -& 2 \cos({g_1} + {g_2}) - \sin^2({g_1} + {g_2}) \Big]
\end{eqnarray}
which naturally reduces to Eq.(\ref{b1}) at $g_{1} = g_{2}$. Interestingly the maximum value of $K^{v}_{13}$ in Eq.(\ref{b2}) is $1.91$ for $g_{1} =0.98 $ and $g_{2} = 1.85$. Thus, instead of using $g_{1}=g_{2}=g$, the consideration of two different intermediate couplings improved the quantum violation of LGI $K_{13}$. However, if the maximum quantum value of $K_{23}$ and $K_{12}$ is calculated by using  $\xi = 1$ and $g_1 = g_2 =g$, we have $(K^{v}_{23b})_{max} = 1$ and $(K^{v}_{12b})_{max} = 1$. Then, no violation of LGIs given by Eq.$(\ref{lgi2})$ and $(\ref{lgi3})$ can be obtained in this case. We demonstrate that for a suitable choice of $g_{1}$ and $g_{2}$ even for $\xi = 1$, the maximum values of $K^{v}_{23}$ and $K^{v}_{12}$ can be largely improved as given in Table I.

Let us now examine if suitable choice of $\xi$ can further increases the quantum violation of LGIs. The QM expressions $K^{v}_{13}, K^{v}_{23}$ and $K^{v}_{12}$ for arbitrary $\xi, g_{1}$ and $g_{2}$ are given in Eqs. (\ref{k13v}), (\ref{k23v}) and (\ref{k12v}) respectively. It can be seen from Table. II that, $K^{v}_{23} = 2 \geq K^{v}_{23b}$ when $\xi = \frac{1}{\sqrt{2}}$. Thus, the choice of von Neumann basis plays an important role. However, Budroni and Emary \cite{budroni14} found the quantum violation of LGI 2.21 by using SDP for all possible choices of basis. Here we wanted to explicitly show that the violation of LGI is basis dependent if von Neumann measurement rule is taken for state reduction. It may also be possible that for a particular value of $\xi$ no violation of any of the LGIs occurs but for the same state and same observable the violation can be considerably large for different value of $\xi$. Note that, for the above choices of state and observables, the value of $K_{12}^{v}$ is lower than L$\ddot{u}$ders bound for any choices of $\xi$.

\begin{table}[ht]
\begin{center}
\begin{tabular} {|p{1.5cm}|p{1.5cm}|p{1.5cm}|p{1.5cm} |p{1.4cm}|}  
\hline
\multicolumn{5}{|c|}{\centering Quantum violations of various LGIs } \\
\hline
\centering LGI & \centering Max. Value & \centering $g_1$ & \centering $g_2$ & \ \ \ \ $\xi$ \\
\hline
\centering $K^{v}_{13}$ & \centering $1.91$ & \centering $0.98$ & \centering $1.85$ & \ \ \ \ $1$ \\
\hline
\centering $ K^{v}_{23}$ & \centering $2.0$ & \centering $\pi$ &  \centering $\pi$ & \ \ \ \  $\frac{1}{\sqrt{2}}$ \\
\hline
\centering $K^{v}_{12}$ & \centering $1.44$ & \centering $2.41$ & \centering $-0.73$ & \ \ \ \ $1$ \\
\hline
\end{tabular}
\label{tab2}
\end{center}
\caption{The quantities $K^{v}_{13}$, $K^{v}_{23}$ and $K^{v}_{12}$ are the quantum values of LGIs expression when von Neumann measurement is performed for different values of $\xi$, $g_{1}$ and $g_{2}$.}
\end{table}

\begin{figure}[ht]
\begin{minipage}[c]{0.5\textwidth}
\includegraphics[width=1\linewidth]{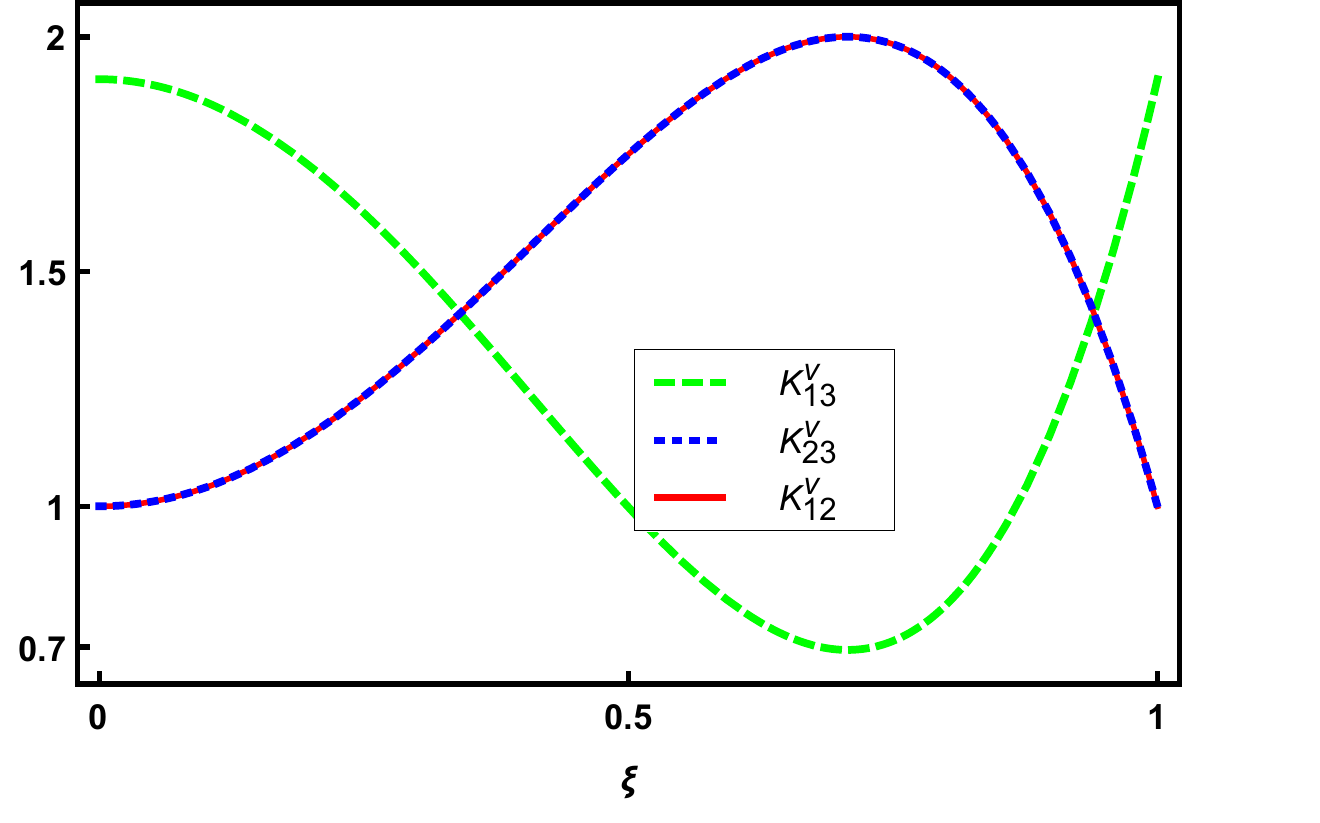}
\end{minipage}\hfill
\begin{minipage}[hc]{0.5\textwidth}
\caption{(color online): The quantum mechanical expressions of $K^{v}_{13}$ and  $K^{v}_{23}$ (calculated for $|\psi_{t_1}\rangle=(0,0,1)^T$) and $K^{v}_{12}$  (calculated for $|\psi_{t_1}\rangle=(1,0,0)^T$) given by (\ref{k13v}), (\ref{k23v}) and (\ref{k12v2}) respectively are plotted as a function of $\xi$ for fixed values of $g_1$ and $g_2$, as given in Table II. }
\label{fig:02}
\end{minipage}
\end{figure}

Next, in order to showing the effect of basis choice for the violation of L$\ddot{u}$ders  bound of $K_{12}$ we consider a state $|\psi_{t_1}\rangle=(1,0,0)^T$ and calculate $K^{v}_{12}$ for that state. The  quantum expressions of $K^{v}_{12}$ is given in  Eq.($\ref{k12v2}$) of the Appendix. The quantum expressions of $K^{v}_{13}$, $K^{v}_{23}$ and $K^{v}_{12}$ given by Eqs. (\ref{k13v}), (\ref{k23v}) and (\ref{k12v2}) respectively are plotted in Fig.{\ref{fig:02}}. We found that $(K^{v}_{12})_{max} = 2$ for $g_1 = g_{2}=\pi $, when $\xi =\frac{1}{\sqrt{2}}$. More suitable choice of state, observables and basis can provide the result obtained in \cite{budroni14} using SDP. The important point we wanted to explicitly pointed out here that the choice of basis plays crucial role in von Neumann measurement scenario and the suitable choice of which can violate the L$\ddot{u}$ders  bound  of LGIs by a considerably large amount. 

In the next section we demonstrate the violation of L$\ddot{u}$ders  bound of non-contextual inequalities (NCIs) and discuss the subtleties involved regarding von Neumann projection rule and discuss its relevance in the violation of realist inequalities.

\section{Violation of L$\ddot{u}$ders  bound of NCIs}

Let  $\hat{A_{1}}$, $\hat{A_{2}}$ and $\hat{A_{3}}$ be the three mutually commuting dichotomic observables. Given a quantum state, the measurement statistics of $\hat{A_{1}}$ is independent of whether measuring with $\hat{A_{2}}$ or $\hat{A_{3}}$.  Similar arguments holds good for $\hat{A_{2}}$ and $\hat{A_{3}}$. This feature is the non-contextuality in QM. In an non-contextual realist model, it is assumed that the individual measured values say, $v(A_{1})$ (fixed by a hidden variable) follows the same context independence as in QM. So that, $v(A_{1})$ is independent of $v(A_{2})$ or $v(A_{3})$. We can then write that in a non-contextual realist hidden variable model the following three inequalities are satisfied. 
\begin{align}
\label{nc1}
  \beta_{31}=\langle \hat{A_1}\hat{A_2}\rangle +\langle \hat{A_2} \hat{A_3}\rangle-\langle \hat{A_3} \hat{A_1}\rangle  \leq 1,
\end{align}
\begin{align}
\label{nc2}
  \beta_{23}=\langle \hat{A_1} \hat{A_2}\rangle -\langle \hat{A_2} \hat{A_3}\rangle+\langle \hat{A_3} \hat{A_1}\rangle  \leq 1,
\end{align}
\begin{align}
\label{nc3}
 \beta_{12} = -\langle \hat{A_1} \hat{A_2}\rangle +\langle \hat{A_2} \hat{A_3}\rangle+\langle \hat{A_3} \hat{A_1}\rangle  \leq 1,
\end{align}

It is evident that none of the above inequalities will be violated by QM, if the L$\ddot{u}$ders rule is used. This is due to the fact that the triple-wise joint probability $P(A_1,A_2, A_3)$ exists whose suitable marginal correctly provides all pair-wise joint probabilities satisfied by QM. Thus, the L$\ddot{u}$ders  bound of those NCIs is also 1. In their paper \cite{budroni14},  Budroni and Emary remarked that the L$\ddot{u}$ders  bound of NCIs can not be violated by using von Neumann rule. We first demonstrate that their inference is not correct and then discuss the suitabilities involved in the von Neumann rule. 

For this, let us consider the observables $\hat{A_{i}}=I-2|\alpha_i\rangle\langle\alpha_i|$ with $\langle\alpha_i|\alpha_j\rangle=\delta_{ij}$  where $i,j=1,2,3$. We take specific examples where $|\alpha_{1}\rangle=(-1,0,1)^{T}/\sqrt{2}$, $|\alpha_{2}\rangle=(1,0,1)^{T}/\sqrt{2}$ and $|\alpha_{3}\rangle=(0,1,0)^{T}$. Eigenstates  of $\hat{A_{1}}$ may be written as  $|a_{1}\rangle=(-1,0,1)^{T}/\sqrt{2}$, $|a_{2}\rangle=(1,0,1)^{T}/\sqrt{2}$ and $|a_{3}\rangle=(0,1,0)^{T}$ with eigenvalues $(-1,1,1)$ respectively. Similarly, the eigenstates of $\hat{A_{2}}$ can be written as  $|b_{1}\rangle=(1,0,1)^{T}/\sqrt{2}$, $|b_{2}\rangle=(-1,0,1)^{T}/\sqrt{2}$ and $|b_{3}\rangle=(0,1,0)^{T}$ with eigenvalues $(-1,1,1)$ respectively, and the eigenstates of $\hat{A_{3}}$ are  $|c_{1}\rangle=(0,1,0)^{T}$, $|c_{2}\rangle=(0,0,1)^{T}$ and $|c_{3}\rangle=(1,0,0)^{T}$ with eigenvalues $(-1,1,1)$ respectively. 

As we have already pointed out that the choices of basis for invoking the von Neumann projection rule is not unique. For example, for observable $\hat{A_{1}}$, the eigenstates can be chosen as $|a_{1}\rangle=(-1,0,1)^{T}/\sqrt{2}$, $|a'_{2}\rangle=\epsilon |a_{2}\rangle + \sqrt{1-\epsilon^{2}}|a_{3}\rangle$ and $|a'_{3}\rangle=\sqrt{1-\epsilon^{2}} |a_{2}\rangle - \epsilon |a_{3}\rangle$ with eigenvalues $(-1,1,1)$ respectively.  Similarly, one can choose the eigenstates of $\hat{A_{2}}$ are $|b_{1}\rangle=(1,0,1)^{T}/\sqrt{2}$, $|b'_{2}\rangle=\lambda |b_{2}\rangle + \sqrt{1-\lambda^{2}}|b_{3}\rangle$ and $|b'_{3}\rangle=\sqrt{1-\lambda^{2}} |b_{2}\rangle - \lambda |b_{3}\rangle$ with eigenvalues $(-1,1,1)$ respectively, and the eigenstate of $\hat{A_{3}}$ are  $|c_{1}\rangle=(0,1,0)^{T}$, $|c'_{2}\rangle=\delta |c_{2}\rangle + \sqrt{1-\delta^{2}}|c_{3\rangle}$ and $|c'_{3}\rangle=\sqrt{1-\delta^{2}} |c_{2}\rangle - \delta |c_{3}\rangle$ with eigenvalues $(-1,1,1)$ respectively. Then there are infinite number of basis choice for implementing von Neumann measurement for different choices of $\epsilon, \lambda, \delta \in [0,1]$.

We examine how different values of $\epsilon$, $\lambda$ and $\delta$ play important role for improving the violaton of different NCIs. By considering the state 
$$|\psi \rangle=(\sin (\theta ) \sin (\phi ),\cos (\theta ) \sin (\phi ),\cos (\phi ))^{T}$$
the quantum expression $\beta ^{l}_{31}$, $\beta ^{l}_{23}$ and $\beta ^{l}_{12}$ are calculated by using L$\ddot{u}$ders projection rule is given by
\begin{align}
\label{nc1L}
\beta ^{l}_{31}=1 - 2 \Big[\cos ( \phi ) + \sin (\theta ) \sin (\phi )\Big]^2
\end{align}
\begin{align}
\label{nc2L}
\beta^{l}_{23}= 1- 2 \Big[\cos ( \phi ) - \sin (\theta ) \sin (\phi )\Big]^2
\end{align}
and
\begin{align}
\label{nc3L}
\beta ^{l}_{12}=1 -4 \cos^2 (\theta ) \sin ^2(\phi )
\end{align}
It is then straightforward to see that the maximum values of $\beta ^{l}_{31}$, $\beta ^{l}_{23}$ and $\beta ^{l}_{12}$ cannot exceed $1$ for any possible choices of $\theta$ and $\phi$, as expected.
 
\begin{figure}[ht]
\begin{minipage}[c]{0.5\textwidth}
\includegraphics[width=1\linewidth]{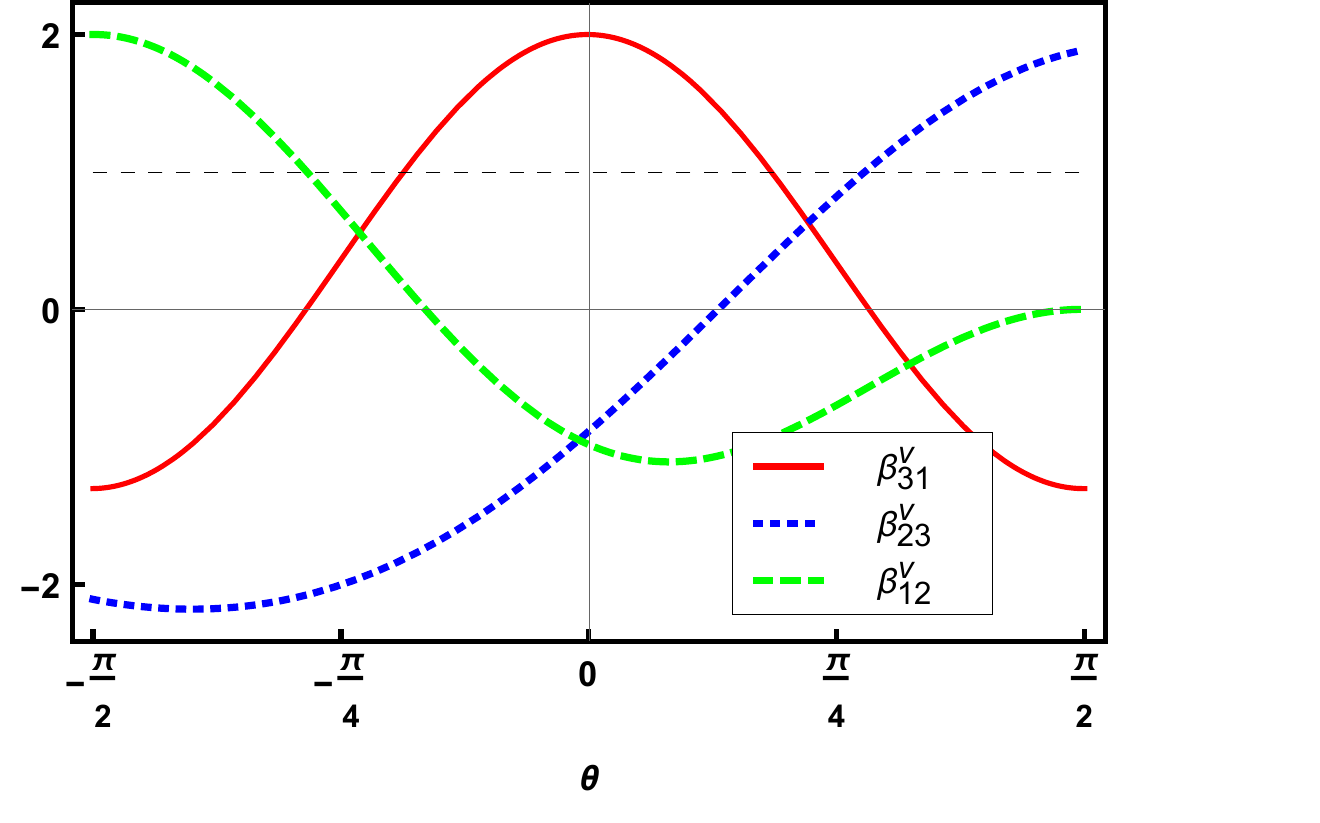}
\end{minipage}\hfill
\begin{minipage}[hc]{0.5\textwidth}
\caption{(color online).  The quantum mechanical expressions $\beta ^{v}_{31}$, $ \beta ^{v}_{23}$ and $ \beta ^{v}_{12}$ given by Eq. (\ref{n13v}), (\ref{n23v}) and (\ref{n12v}) respectively are plotted as a function of $\theta$ for fixed values of $\phi, \epsilon, \lambda$ and $\delta$ given by Table IV.}
\label{fig:04}
\end{minipage}
\end{figure}

The quantum expressions $\beta ^{v}_{31}$, $\beta ^{v}_{23}$ and $\beta ^{v}_{12}$ for von Neumann rule are given by Eq.(\ref{n13v}),(\ref{n23v}) and (\ref{n12v}) respectively in the Appendix A. We maximize $\beta ^{v}_{31}$, $\beta ^{v}_{23}$ and $\beta ^{v}_{12}$ by suitably choosing the relevant parameter $\theta, \phi, \lambda$ and $\delta$ as given in Table. IV. In Fig.\ref{fig:04}, we plotted $\beta ^{v}_{31}$, $ \beta ^{v}_{23}$ and $ \beta ^{v}_{12}$ as a function of $\theta$ with fixed values of $\phi,\epsilon,\lambda$ and $\delta$ as given in the Table. IV. We found that quantum values of all $\beta ^{v}_{31}$, $\beta ^{v}_{23}$ and $\beta^{v}_{12}$ reach $2$, exceeding L$\ddot{u}$ders bound $1$.
\begin{table}
\begin{center}
\begin{tabular} {|p{1.1cm}|p{1.1cm}| p{1.1cm} |p{1.1cm}|p{1.1cm}| p{1.1cm} |p{1.1cm}|} 
\hline
\multicolumn{7}{|c|}{Quantum violations of NCIs } \\
\hline
\centering     
NCI & Max. value & \  \ $\phi$ & \  \ $\theta$ & \ \ $\epsilon$ & \  \ $\lambda$ & \  \ $\delta$  \\
\hline
\ \ $\beta^{v}_{31}$ & \ \ $2$ & \ \ $\pi/2$ & \ \ $0$ & \ \ $0$ & \ \ $0.1$ & \ \ $0.7$ \\
\hline		
\ \ $\beta^{v}_{23}$ & \ \ $2$ & \ \ $\pi/4$ & \ \ $\pi/2$  & \ \ $0.7$ & \ \ $1$ & \ \ $0.7$  \\
\hline
\ \ $\beta^{v}_{12}$ & \ \ $2$ & \ \ $3\pi/4$ & \ \ $\pi/2$  & \ \ $1$ & \ \ $1$ & \ \ $1$  \\
\hline
\end{tabular}
\label{tab4}
\caption{The quantities $\beta ^{v}_{31}$, $ \beta ^{v}_{23}$ and $ \beta ^{v}_{12}$ denote the quantum values of different NCIs using von Neumann projection rule.}
\end{center}
\end{table}

We thus exhibited the violation of L$\ddot{u}$ders  bound of NCIs  even when the observables $\hat{A_{1}}$, $\hat{A_{2}}$ and $\hat{A_{3}}$ are mutually commuting. This is in contrast to the claim in \cite{budroni14}. Again, the different choice of von Neumann basis provides different violation of L$\ddot{u}$ders  bound of NCIs. Conceptually, this result is very surprising. Since the NCIs given by Eqs.$(\ref{nc1}-\ref{nc3})$ are derived by assuming that the existence of the joint probability distribution $P(A_1,A_2, A_3)$ in a non-contextual theory. The quantum violation of NCIs can only be obtained when $P(A_1,A_2, A_3)$ does not exist in QM. Since $\hat{A_{1}}$, $\hat{A_{2}}$ and $\hat{A_{3}}$ are mutually commuting then triple-wise joint probability exists in QM too. Hence, as we have already mentioned in introductory section, a kind of non-classicality is introduced through the von Neumann rule which enables the quantum violation of NCIs given in Eqs.(\ref{nc1}-\ref{nc3}). In the next section, we provide a detail discussions regarding the meaning of such violation and its relevance with respect to the usual violation of realist inequalities.

\section{Summary and Discussion}
In this paper we questioned the conceptual relevance of the violation of L$\ddot{u}$ders bound of LGIs and NCIs obtained through the degeneracy breaking projective rule for state reduction. The improved violations of realist inequalities are the signature of more non-classicality and  any such improvement is useful for testing the concerned inequalities experimentally.  For a dichotomic system the L$\ddot{u}$ders bound of three time LGIs is restricted to $1.5$. Budruni and Emary \cite{budroni14} have shown that for dichotomic observables in qutrit system the use of von Neumann measurement projection rule for state reduction can provide quantum violation of LGI that exceeds the L$\ddot{u}$ders bound $1.5$. However, they have considered a particular choice of the von Neumann basis. We have pointed out that the choice of basis for implementing the so called von Neumann rule for state reduction is not unique and there can be infinitely many choices possible. We showed that the violation of L$\ddot{u}$ders bound of LGIs is dependent on the choices of von neumann basis. 

LGIs are pertaining to the sequential measurement of two non-commuting observables. In contrast, in contextuality test the pair-wise joint measurements are assumed for commuting observables. The NCIs we have considered here involve three commuting observables. Since triple-wise joint probability  distribution of those commuting observables exist in QM, no violation of NCIs is expected if L$\ddot{u}$ders rule is used. Budroni and Emary \cite{budroni14} claimed that this feature is true even if one uses the von Neumann projection rule for state reduction. We showed here that their inference is not correct and suitable choice of von Neumann basis can provide a considerably large  violation of L$\ddot{u}$ders  bound of NCIs. One may find this result is particularly interesting for experimental testing as compared to LGI test because in the non-contextuality test one does not require to guarantee the non-invasiveness condition of the measurement. However, the meaning of such violation remains questionable. 

Let us  again return to the issue regarding the implications of such violations of L$\ddot{u}$ders bound of LGIs and NCIs. For this, we consider the NCI given by Eq.(\ref{nc1}). As already mentioned, for implementing von Neumann rule in a sequential measurement of $\langle A_{1}A_{2}\rangle$ one has to measure another observable $A_{1}^{\prime}$, prior to $A_1$. Note that, the measurement of $A_{1}^{\prime}$ is non-detective, i.e., only the state reduction due to the measurement of $A_{1}^{\prime}$ is taken into account. Similarly,  an observable $\langle A_{2}A_{3}\rangle$ requires $A_{2}^{\prime}$ to be measured before $A_{2}$. Moreover, the different choice of von Neumann basis requires different $A_{1}^{\prime}$ and similarly different $A_{2}^{\prime}$.  There is whatsoever no reason to consider $A_{1}^{\prime}$ and $A_{2}^{\prime}$ are same in general. Then the initial density matrices for the sequential measurements of  $\langle A_{1}A_{2}\rangle$ and $\langle A_{2}A_{3}\rangle$ can be different. One can suitably choose $A_{1}^{\prime}$ and $A_{2}^{\prime}$ to obtain an improved violation of NCI exceeding the L$\ddot{u}$ders bound. But, neither the observables $A_{1}^{\prime}$ and $A_{2}^{\prime}$  nor the state change information due to the measurement of them was included in deriving the realist bound of NCI given by Eq.(\ref{nc1}). Similar argument holds good for other NCIs and for LGIs. Thus, the violations of L$\ddot{u}$ders bound of NCIs and LGIs has no bearing on the issue of quantum violations of the realist inequalities.

Note however that, the results provided here can be calculated within the framework of standard QM by suitably choosing the prior measurements and can then be experimentally tested. It would then be appealing  to study how such empirically verifiable statistics can be reproduced in an ontological model and would be interesting to see what additional constraint is needed to be imposed in such a model in order to be consistent with the QM.  This calls for further study.

\section*{Acknowledgments}
MQ  acknowledge the Junior Research Fellowship from SERB project (ECR/2015/00026). AKP acknowledges the support from Ramanujan Fellowship research grant (SB/S2/RJN-083/2014).

\begin{widetext}
\appendix
\section{}
The quantum mechanical expressions of $K_{13}, K_{23}$ and $K_{12}$ given by Eq. (\ref{lgi1}), (\ref{lgi2}) and (\ref{lgi3}) respectively are calculated for initial state $|\psi\rangle = (0,0,1)^T$ with different intermediate evolutions between measurements are given by
\begin{eqnarray}
\label{k13v}
K^{v}_{13} &=& \nonumber 1+ \Big(-2 -11\xi^{2} + 11\xi^{4} + 9\xi^{2}(-1 + \xi^{2})\cos(g_{2}) + \cos(g_{1})\big(4 - 21\xi^{2} + 21\xi^{4} + (2 - 15\xi^{2} + 15\xi^{2})\cos(g_{2}) \big) \Big)\\ & & \sin^{2}(\frac{g_{1}}{2})\sin^{2}(\frac{g_{1}}{2})+  (1 - 2\xi^{2} + 2\xi^{4})\sin(g_{1})\sin(g_{2}) - \frac{1}{4}(1 - 6\xi^{2} + 6\xi^{4})\sin(2g_{1})\sin(2g_{2}) 
\end{eqnarray}
\begin{eqnarray}
\label{k23v}
K^{v}_{23} &=& \nonumber\frac{1}{4} \Big[ 4+ \Big(-4 - 3\xi^{2} +3\xi^{4} - 4\cos(g_{2}) + \xi^{2}(-1 + \xi^{2})(4\cos(g_{2}) + 9\cos(2g_{2})) + \cos(g_{1})\big(2 - 21\xi^{2} + 21\xi^{4} +4(1 - 3\xi^{2} \\\nonumber &+& 3\xi^{4})\cos(g_{2}) + (2 - 15\xi^{2} + 15\xi^{4})\cos(2g_{2}) \big) \Big)\sin^{2}(\frac{g_{1}}{2}) - 4 (1 - 2\xi^{2}+ 2\xi^{2})\sin(g_{1})\sin(g_{2})+ (1 - 6\xi^{2} + 6 \xi^{4})\\ & & \sin(2g_{1})\sin(2g_{2}) \Big]
\end{eqnarray}
\begin{eqnarray}
\label{k12v}
K^{v}_{12} &=& \nonumber \frac{1}{4} \Big[4+\Big(-8 -\xi^{2} + \xi^{4} + (-2 -3\xi^{2} + 3\xi^{4})\cos(g_{2}) + \cos(2g_{1})\big(4 - 27\xi^{2} + 27\xi^{4} + (6 - 33\xi^{2} + 33\xi^{4})\cos(g_{2}) \big) \Big)\sin^{2}(\frac{g_{2}}{2})\\\nonumber &-& 8 (1 - 3\xi^{2} +3\xi^{4})\cos(g_{1})\sin^{2}(\frac{g_{2}}{2}) - 4\sin(g_{1})\sin(g_{2}) + 8\xi^{2}(-1 + \xi^{2})(-1 + 3\cos(g_{1})\cos(g_{2}))\sin(g_{1})\sin(g_{2})\\ &+& \sin(2g_{1})\sin(2g_{2}) \Big]
\end{eqnarray}

The quantum mechanical expressions of $K^{v}_{12}$ obtained for the state $|\psi_{t_1}\rangle=(1,0,0)^T$ is given by
 
\begin{eqnarray}
\label{k12v2}
K^{v}_{12} &=&\nonumber\frac{1}{16} \bigg[\xi ^2-\xi ^4+2-16 \cos (g_1) \sin ^2\left(\frac{g_2}{2}\right) \left(\left(3 \xi ^4-3 \xi ^2+1\right) \cos (g_2)+\xi ^4-\xi ^2+3\right)+2 \cos (2 g_2)\nonumber\\&+&4 \cos (2 g_1) \sin ^2\left(\frac{g_2}{2}\right) \left(\left(9 \xi ^4-9 \xi ^2-2\right) \cos (g_2)+3 \xi ^4-3 \xi ^2-4\right)-16 \sin (g_1) \sin (g_2)\nonumber\\&-&4 \sin (2 g_1) \sin (2 g_2)-3 \left(\xi ^2-1\right) \xi ^2 \cos (2 g_2)+4 \left(\xi ^4-\xi ^2+3\right) \cos (g_2)\bigg]
\end{eqnarray}

The quantum mechanical expressions of $\beta_{31}, \beta_{23}$ and $\beta_{12}$ given by Eq.(\ref{nc1}), (\ref{nc2}) and (\ref{nc3}) respectively are calculated for the initial state $|\psi \rangle=(\sin (\theta ) \sin (\phi ),\cos (\theta ) \sin (\phi ),\cos (\phi ))^{T}$. The expressions are given by
\begin{eqnarray}
\label{n13v}
{\beta}^{v}_{31} &=& \nonumber \Big(-1 +2\epsilon^{2} - 2\epsilon^{4} + (-1 + \lambda^{2}) 2\lambda^{2} + 2\delta(1 - 2\delta^{2})\sqrt{1-\delta^{2}} \Big)\cos^{2}(\phi) + \Big((\epsilon - \lambda)(\epsilon + \lambda)(-1 + \epsilon^{2} + \lambda^{2}) + (1 - 3\epsilon^{2} \\\nonumber &+& 3\epsilon^{4} +  3\lambda^{2} - 3\lambda^{4})\cos(2\theta) + 2\delta(-1 + 2\delta^{2})\sqrt{1 - \delta^{2}}\sin^{2}(\theta) + \sqrt{2}\big(\epsilon(1 - 2\epsilon^{2})\sqrt{1-\epsilon^{2}} + \lambda(1 - 2\lambda^{2})\sqrt{1-\lambda^{2}}\big)\sin(2\theta) \Big)\sin^{2}(\phi)\\\nonumber  &-&  \Big(\sqrt{2}\big(\epsilon(-1 + 2\epsilon^{2})\sqrt{1-\epsilon^{2}} + \lambda(-1 + 2\lambda^{2})\sqrt{1 - \lambda^{2}}\big)\cos(\theta) + \sin(\theta) + 2 (- \epsilon^{2} + \epsilon^{4}- \lambda^{2} +\lambda^{4} + 2\delta^{2} - 2\delta^{4})\sin(\theta) \Big)\sin(2\phi) 
\\
\end{eqnarray}

\begin{eqnarray}
\label{n23v}
{\beta}^{v}_{23} &=& \nonumber \big(-1 +2\epsilon^{2} + 2\epsilon^{4} + 2\lambda^{2} - 2\lambda^{4} + 2\delta(-1 + 2\delta^{2})\sqrt{1-\delta^{2}} \big)\cos^{2}(\phi) + \Big(-\epsilon^{2} + \epsilon^{4} - \lambda^{2} + \lambda^{4} + (1 - 3\epsilon^{2} + 3\epsilon^{4} - 3\lambda^{2}\\\nonumber &+& 3\lambda^{4})\cos(2\theta) + 2\delta(1 - 2\delta^{2})\sqrt{1 - \delta^{2}}\sin^{2}(\theta) + \sqrt{2}\big(\epsilon(1 - 2\epsilon^{2})\sqrt{1-\epsilon^{2}} + \lambda(-1 + 2\lambda^{2})\sqrt{1-\lambda^{2}}\big)\sin(2\theta) \Big)\sin^{2}(\phi)\\ \nonumber &+&  \Big(\sqrt{2}\big(\epsilon(1 - 2\epsilon^{2})\sqrt{1-\epsilon^{2}} + \lambda(1 - 2\lambda^{2})\sqrt{1 - \lambda^{2}}\big)\cos(\theta) + \sin(\theta) + 2 (\epsilon^{2} - \epsilon^{4}- \lambda^{2} +\lambda^{4} + 2\delta^{2} - 2\delta^{4})\sin(\theta) \Big)\sin(2\phi)  
\\
\end{eqnarray}

\begin{eqnarray}
\label{n12v}
{\beta}^{v}_{12} &=& \nonumber \big(-1 +2\epsilon^{2} + 2\epsilon^{4} + (-1 + \lambda^{2}) 2\lambda^{2} + 2\delta(-1 + 2\delta^{2})\sqrt{1-\delta^{2}} \big)\cos^{2}(\phi) - \Big(1 - \epsilon^{2} + \epsilon^{4} - \lambda^{2} + \lambda^{4}+ (2 - 3\epsilon^{2} + 3\epsilon^{4}\\\nonumber &-& 3\lambda^{2} + 3\lambda^{4})\cos(2\theta) + 2\delta(-1 + 2\delta^{2})\sqrt{1 - \delta^{2}}\sin^{2}(\theta) + \sqrt{2}\big(\epsilon(1 - 2\epsilon^{2})\sqrt{1-\epsilon^{2}} + \lambda(-1 + 2\lambda^{2})\sqrt{1-\lambda^{2}}\big)\sin(2\theta) \Big)\sin^{2}(\phi)\\\nonumber &+&  \Big(\sqrt{2}\big(\epsilon(-1 + 2\epsilon^{2})\sqrt{1-\epsilon^{2}} + \lambda(-1 + 2\lambda^{2})\sqrt{1 - \lambda^{2}}\big)\cos(\theta) + (-1 -  2\epsilon^{2} + 2\epsilon^{4} + 2\lambda^{2} - 2\lambda^{4} + 4\delta^{2} - 4\delta^{4})\sin(\theta) \Big)\sin(2\phi)
\\
\end{eqnarray}

\end{widetext}


\begin{thebibliography}{99}
\bibitem{epr35} A. Einstein, B. Podolsky, and N. Rosen, Phys. Rev. \textbf{47}, 777 (1935).
\bibitem{bell64} J. S. Bell, Physics \textbf{1}, 195 (1964).
\bibitem{genovese} M. Genovese,  Phys. Rep. \textbf{413}, 319 (2005).
\bibitem{rev} N. Brunner \emph{et al}., Rev. Mod. Phys. \textbf{86}, 419 (2014).

\bibitem{ks67} S. Kochen and E. Specker, J. Math. Mech.  \textbf{17}, 59 (1967).
\bibitem{peres} A. Peres, Phys. Lett. A \textbf{151}, 107 (1990).
\bibitem{mermin90} N. D. Mermin, Phys. Rev. Lett. \textbf{65}, 3373 (1990); Rev. Mod. Phys. \textbf{65}, 803 (1993).
\bibitem{cab08} A. Cabello, Phys. Rev. Lett. \textbf{101}, 210401 (2008).
\bibitem{pan10} A. K. Pan, EPL \textbf{90}, 40002 (2010).
\bibitem{theocontext} A. Cabello, and G. Garcia-Alcaine,  J. Phys. A \textbf{29}, 1025 (1996); R. W. Spekkens, Phys. Rev. A \textbf{71}, 052108 (2005); A. A. Klyachko \emph{et al.}, Phys. Rev. Lett. \textbf{101}, 020403 (2008); A. K. Pan and D. Home,  Phys. Lett. A \textbf{373}, 3430(2009); S. Yu and C. H. Oh,  Phys. Rev. Lett. \textbf{108}, 030402 (2012); R. Kunjwal and R. W. Spekkens, Phys. Rev. Lett. \textbf{115}, 110403 (2015). 
\bibitem{kle} M. Kleinmann \emph{et al}., Phys. Rev. Lett. \textbf{109}, 250402 (2012).
\bibitem{expcontext}  M. Michler, H. Weinfurter and M. Zukowski, Phys. Rev. Lett. \textbf{84}, 5457 (2000); Y. Hasegawa \emph{et al.}, Nature \textbf{425}, 45 (2003); G. Kirchmair \emph{et al}.,  Nature, \textbf{460}, 494 (2009); E. Amselem \emph{et al}., Phys. Rev. Lett. \textbf{103}, 160405 (2009); O. Moussa \emph{et al}., Phys. Rev. Lett. \textbf{104}, 160501 (2010); R. Lapkiewicz \emph{et al.}, Nature \textbf{474}, 490 (2011); M. D. Mazurek, \emph{et al.}, Nat. Commun. \textbf{7}, 11780 (2016).

\bibitem{lg85} A. J. Leggett and A. Garg, Phys. Rev. Lett. \textbf{54}, 857 (1985).
\bibitem{lagtt2002} A. J. Leggett, J. Phys.: Condens. Matter \textbf{14}, R415 (2002).

\bibitem{emary12} C. Emary, N. Lambert, and F. Nori, Phys. Rev. B \textbf{86}, 235447
(2012).


\bibitem{maroney14} O. J. E. Maroney and C. G Timpson, arxiv: 1412.613v1.
\bibitem{kofler13} J. Kofler and C. Brukner, Phys. Rev. A \textbf{87}, 052115 (2013).
\bibitem{budroni15} C. Budroni \emph{et al}.,  Phys. Rev. Lett. \textbf{115}, 200403 (2015).
\bibitem{saha15} D. Saha \emph{et al}., Phys. Rev. A \textbf{91}, 032117 (2015).

\bibitem{halliwell16} J. J. Halliwell, Phys. Rev. A \textbf{93}, 022123
(2016).

\bibitem{budroni14} C. Budroni and C. Emary,  Phys. Rev.
Lett. \textbf{113}, 050401 (2014).
\bibitem{swati17} S. Kumari and A. K. Pan, Euro. Phys. Lett. \textbf{118}, 50002 (2017).
\bibitem{pan17} S. Kumari and A. K. Pan, Phys. Rev. A \textbf{96}, 042107 (2017).

\bibitem{kofler08} J. Kofler and C. Brukner, Phys. Rev. Lett. \textbf{101}, 090403 (2008).
\bibitem{emary} C. Emary, N. Lambert and F. Nori,  Rep. Prog. Phys. \textbf{77}, 016001 (2014).


\bibitem{lambert} N.Lambert \emph{et al}.,  Phys. Rev. A \textbf{94}, 012105 (2016).

\bibitem{goggin11} M. E. Goggin \emph{et al.}  Proc. Natl. Acad. Sci. U.S.A. \textbf{108}, 1256 (2011).
\bibitem{knee12} G. C. Knee \emph{et al}., Nat. Commun. \textbf{3}, 606 (2012).
\bibitem{laloy10} A. Palacios-Laloy \emph{et al}., Nat. Phys. \textbf{6}, 442 (2010).
\bibitem{george13} R. E. George \emph{et al}., Proc. Natl. Acad. Sci. U.S.A. \textbf{110}, 3777 (2013).


\bibitem{knee16} G. C. Knee \emph{et al}., Nat. Commun. \textbf{7}, 13253 (2016).
\bibitem{kati1} H. Katiyar \emph{et al}., N. J. Phys. \textbf{19}, 023033 (2017).
\bibitem{wang02} K. Wang \emph{et al}., arXiv:1701.02454 (2017).




\bibitem{chsh69} J. F. Clauser, M. A. Horne, A. Shimony and R. A. Holt,  Phys. Rev. Lett. \textbf{23}, 880, (1969).
\bibitem{cri} B. C. Cirelson,  Lett. Math. Phys. \textbf{4}, 93 (1980).
\bibitem{lu} G. L$\ddot{u}$ders,  Ann. Phys. (Leipzig) \textbf{6}, 322 (1951).
\bibitem{budroni13} C. Budroni \emph{et al}., Phys. Rev. Lett. \textbf{111}, 020403 (2013).
\bibitem{heger} G. C. Hegerfeldt and R. Sala Mayato, Phy Lett. A \textbf{375}, 3167 (2011).

\bibitem{pan} A. K. Pan and K. Mandal, Int. J. Theor. Phys. \textbf{55}, 3472 (2016).

\end{thebibliography}
\end{document}